# Guided assembly of cellular network models from knowledge in literature

Yasmine Ahmed[1] and Natasa Miskov-Zivanov[1,2,*]

*Abstract*— Computational modeling is crucial for understanding and analyzing complex systems. In biology, model creation is a human dependent task that requires reading hundreds of papers and conducting wet lab experiments, which would take days or months. To overcome this hurdle, we propose a novel automated method, that utilizes the knowledge published in literature to suggest model extensions by selecting most relevant and useful information in few seconds. In particular, our novel approach organizes the events extracted from the literature as a collaboration graph with additional metric that relies on the event occurrence frequency in literature. Additionally, we show that common graph centrality metrics vary in the assessment of the extracted events. We have demonstrated the reliability of the proposed method using three different selected models, namely, T cell differentiation, T cell large granular lymphocyte, and pancreatic cancer cell. Our proposed method was able to find high percent of the desired new events with an average recall of 82%.

## I. Introduction

Computational mechanistic models are very often used to explain complicated systems [1][2][3]. This helps identify gaps in our understanding, find answers to new questions and identify missing information. In biology, model creation is highly dependent on human intervention. It requires reading hundreds of papers to extract useful information, incorporating background and common-sense knowledge of domain experts, and conducting wet-lab experiments. Moreover, the amount of biological data is constantly growing, further augmenting the issues of data inconsistency and fragmentation. Therefore, automating the process of building new, and extending existing models, is critical for consistent, comprehensive and robust studies of biological systems. Following the questions raised about systems under study, search queries can be defined formally to automatically select published papers with relevant information. Several reading engines have been recently developed, focusing on biomedical literature and automatically extracting hundreds of thousands of events from thousands of papers within hours [4][5]. To add this information to existing models, or to build new models from it, one needs methods and tools for systematic selection of most useful information from this large machine reading output.

The INDRA (Integrated Network and Dynamical Reasoning Assembler) database and tool [6] has been proposed to assemble the biomolecular signaling pathways. This is done by collecting and scoring new information extracted from literature, by natural language processing algorithms, or from pathway databases. To evaluate the confidence in the collected information, each statement is evaluated and has an overall support, a belief score, computed as the joint probability of correctness implied by the evidence. Another tool, FLUTE (FiLter for Understanding True Events)[7], has been recently proposed to further filter the extracted interactions using public databases in order to eliminate incorrect or non-relevant information extracted by machine readers. FLUTE also allows users to enter a threshold for the interaction score when selecting interactions.

In this work, we propose a novel method to automatically assemble models, by selecting most relevant and useful information from published literature. This is achieved by identifying the most influential events in the newly extracted information, and then scoring these events using the occurrence frequency of events and graph centrality metrics. Unlike INDRA and FLUTE, our proposed methodology examines events extracted from literature in the context of a collaboration graph and the measure of the occurrence frequency in literature. We also explore the role of several graph centrality metrics in identifying the most influential events. To this end, we propose a heuristic to determine which centrality metrics are crucial for finding those influential events. Our methodology takes at most a few seconds to execute thousands of in silico experiments, which would otherwise take months, or would be impractical, to conduct in a wet lab. We evaluate our method of model assembly using three benchmark models.

The main contributions of this work are: 1) application of the concept of a collaboration graph in guiding model extension; 2) a metric for event ranking based on their occurrence frequency in literature; 3) a method to evaluate the importance of graph centrality characteristics in finding influential events; 4) application of the proposed methods on several case studies in biology.

## II. Background

### A. Information extraction from biomedical literature

The machine reading of biomedical literature [11][4] extracts *events* between biological *entities* from papers. The events are mechanistic interactions such as post-translational modifications (e.g., binding, phosphorylation, ubiquitination, etc.), transcription, translation, translocation, or more qualitative causal relationships, such as positive and negative influences on the amount or activity of entities. The entities can be components of signaling pathways and gene regulation, such as proteins, genes, and chemicals, or even biological processes.

For each extracted entity, reading engines provide its name, the unique standard identifier (ID) found in public databases (e.g., UniProt [8], GO [9], HMDB [10]), and the entity type. Besides the entity and event information, machine reading also provides the event evidence, the published paper, and the sentence from which the event was extracted.

[1]Electrical and Computer Engineering Department, [2,*]Bioengineering Department, Computational and Systems Biology Department, University of Pittsburgh, Pittsburgh, PA.

In the rest of this paper, we will refer to the set of events extracted by machine readers as *Extracted Event Set* (*EES*). As the same event can be extracted from many different papers, and even multiple times from the same paper, there can be total of *n* events in EES, and *m* distinct events, where ($n \geq m$).

### B. Dynamic network models

Cellular signaling pathways can be modeled as directed graphs $G(V, E)$, with a set of nodes $V$ representing pathway elements, and a set of edges $E$ representing interactions between elements [3]. Each directed edge $e(v_e^s, v_e^t) \in E$ represents a directed interaction in which the source node of the edge, $v_e^s \in V$, is a regulating element, and the target node of the edge, $v_e^t \in V$, is a regulated element. Here, we will refer to the set of all positive and negative regulators of an element (activators and inhibitors, respectively) as its *influence set*. Besides their network (graph-based) structure, these models are also dynamic, they contain update functions that are used to change states of model elements, and thus, enable simulation of model element behavior in time [11][12].

To represent all the details of a model, including its network structure and the update functions, we use the BioRECIPES tabular element-based format proposed in [13], as it is able to capture all the relevant information for dynamic and causal modeling. The BioRECIPES format includes a number of element and influence set attributes, such as name, type (e.g., protein, chemical, gene, biological process), identifier from a database (e.g., UniProt [8]), variable that represents the element state, all regulators in the influence set, and evidence statements with the text from which the event was obtained (when available).

## III. PROPOSED METHODOLOGY

In this section, we describe the main steps of our proposed methodology, which are also outlined in Figure 1.

### A. ECLG creation

Following the notion of a collaboration graph that is often used to model social networks [21], we introduce the *Event Collaboration Graph* (*ECLG*). In the social network domain, nodes represent participants and edges connect two nodes whenever there is a collaborative relationship between them. Similarly, we define the ECLG as an undirected graph $\mathcal{G}(\mathcal{E}, C)$, where $\mathcal{E}$ is a set of graph nodes, each representing a distinct event *e* in EES, $C$ is a set of undirected graph edges, each edge $c(e_i, e_j)$ indicating a co-occurrence in the same paper of its adjacent nodes, $e_i$ and $e_j$ (i.e., the two events represented by these nodes).

### B. Frequency class metric

To measure the frequency of occurrence within EES of individual distinct events that belong to an ECLG, we propose to use a computational linguistic concept for calculating word frequency, called Häufigkeitsklasse or *frequency class* (*FC*) [19][20]. Here, given the EES (with *n* total events and *m* distinct events), we compute the frequency class value, $FC_i$, for each extracted distinct event $e_i$, where $i=1,..,m$ and $\lfloor .. \rfloor$ is the floor function:

$$FC_i = \left\lfloor 0.5 - log_2 \frac{f_i}{f_{max}} \right\rfloor \quad (1)$$

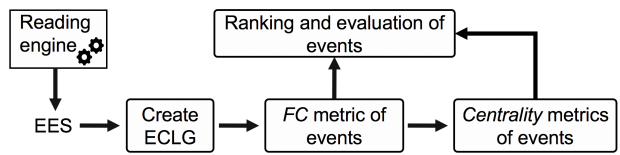

Figure 1. Proposed methodology workflow.

We denote the frequency of each distinct event $e_i$, that is, the overall number of occurrences of event $e_i$ within EES, as $f_i$. We also identify all distinct events for which $f_{max} = \max(\{f_i: i=1,..,m\})$. As can be concluded from equation (1), the most frequent event, that is, any event $e_i$ for which $f_i=f_{max}$, will have $FC_i = 0$, while any event half as frequent as the most frequent event will have $FC_i = 1$ (due to logarithm with base 2).

It is worth mentioning that unlike simple naïve event count, the frequency class-based metric helps group the events within an EES into several categories. This will allow modelers to consider the events within or across these categories. Additionally, setting a threshold based on a simple event count sounds arbitrary and does not account for the occurrence frequency of the other events. In contrast, the frequency class-based metric is computed for each event with respect to the most frequent event, which helps modelers to choose a threshold (as will be discussed in Section III.E) and discard the less frequent events.

### C. Centrality metrics

A number of centrality metrics have been introduced to identify and rank most influential or central nodes in large networks [22][23]. For any given network, the selection of most suitable centrality metrics is affected by the network's topology. Here, we are interested in exploring the correlation between several centrality metrics (defined as follows) and our proposed frequency class metric.

*Degree* (*D*) of a node is the number of all its adjacent edges in an undirected graph. *Neighborhood connectivity* (*NC*) of a node is an average of the *D* value of all its neighbors, where a neighbor is a node connected with the given node via an edge. Nodes with more neighbors tend to have these neighbors more connected (larger *NC*), while nodes with few neighbors usually have their neighbors less connected (smaller *NC*). *Betweenness centrality* (*BC*) of a node is the number of shortest paths between any other couple of nodes, that pass through the given node. High *BC* value for a node indicates that the node, for certain paths, is crucial to maintain node connections. *Closeness centrality* (*CC*) of a node is an inverse of a sum of the lengths of shortest paths between the node and all the other nodes in the graph. Higher *CC* value for a node indicates more proximity to other nodes. *Radiality* (*R*) of a node is computed by first finding the lengths of shortest paths between the node and every other node in the graph, then subtracting the value of the diameter (the maximal possible distance between nodes) from each shortest path length, and finally adding all the resulting values. If a node radiality is high compared to the average radiality of the network, this means that the node is generally closer to the other nodes, however, if the radiality is low, the node is peripheral.

### D. Relationship between centrality and FC metrics

To determine in an automated way which centrality metrics are most correlated with the FC metric for a given network, we use the *permutation feature importance* (PFI), a machine learning technique described in [25]. A typical supervised machine learning problem is composed of: (*i*) a data set $\mathcal{D}$, (*ii*) a set of features $\Phi$, and (*iii*) a corresponding target class $\mathcal{T}$. For the application of the PFI algorithm in this work, the nodes of the ECLG form the data set $\mathcal{D}$, the centrality metric values $D$, $NC$, $BC$, $CC$, and $R$ form the feature set $\Phi$, and the $FC$ metric values previously determined for the ECLG nodes are used as the target class value set $\mathcal{T}$.

Next, we use the PFI algorithm to determine which feature (i.e., which centrality metric) in $\Phi$ contributed the most to the target class value ($FC$) in $\mathcal{T}$ of each data point in $\mathcal{D}$ (node in the ECLG). The class of each data point in $\mathcal{D}$ in a supervised machine learning problem is obtained using a trained classifier, and we use here the *k*-nearest neighbor (KNN) classifier [26]. KNN is considered one of the top ten most effective data mining algorithms for their ability to generate simple but powerful classifiers [27].

The details of the PFI algorithm are the following. For the given data $\mathcal{D}$ (i.e., the ECLG nodes) and each feature (i.e., centrality metric), we determine the corresponding feature vector, i.e., a vector of values for the given feature in each data point. The PFI algorithm then conducts multiple iterations, in each iteration randomly shuffling one feature vector to obtain a corrupted version of data $\mathcal{D}$. For a given feature $\varphi \in \Phi$, and an iteration $l = 1..L$, the algorithm computes a score $s_{\varphi,l}$, which is used to indicate the accuracy of the classifier (how closely it matches the target class). The *importance score* $p_\varphi$ is then computed for each feature $\varphi$ using this equation (2):

$$p_\varphi = s_{\varphi,0} - \frac{1}{L}\sum_{l=1}^{L} s_{\varphi,l} \qquad (2)$$

where $s_{\varphi,0}$ is computed at the beginning of the algorithm, before any shuffling. The PFI algorithm provides as output the importance score of each feature (centrality metric), thus quantifying the contribution of these features to the given classification of the ECLG.

### E. Selection of candidate extension events

The events selected either using the FC metric or the centrality metrics are considered potential candidates for model extension or assembly and are selected as follows.

We rank all the events in the ECLG (i.e., in set $\mathcal{E}$) in ascending order of $FC$ values, i.e., from the most to the least frequent event. Next, we determine a threshold $FC$ value. This threshold can be determined in different ways, for example, it can be provided as a fixed input parameter, or it can be determined based on the used EES. As we will discuss in Section V, for our case studies we consider the threshold to be an average $FC$ value, $FC_{avg}$, computed across all nodes (events) in set $\mathcal{E}$. We then create a new set $\mathcal{E}^{FC}$, a subset of $\mathcal{E}$, including all events from $\mathcal{E}$ with $FC \leq FC_{avg}$. We refer to the events in $\mathcal{E}^{FC}$ as *FC candidate events*. In other words, we remove less frequent events from the original ECLG to form a smaller graph $G^{FC}$ ($\mathcal{E}^{FC}$, $C^{FC}$). This step will effectively remove edges from the original set $C$, thus making $C^{FC}$ a subset of $C$.

Table 1. Summary of the baseline model graph measures and ECLG metric values before and after the removal of less frequent events (ECLG$^{original}$ and ECLG$^{FC}$, respectively) for T cell, TLGL, and PCC use cases.

| Study | T cell | | T-LGL | | PCC | |
|---|---|---|---|---|---|---|
| Model measures | Baseline | Gold | Baseline | Gold | Baseline | Gold |
| number of nodes | 39 | 43 | 41 | 60 | 241 | 257 |
| number of edges | 60 | 74 | 113 | 193 | 280 | 373 |
| EES measures | ECLG$^{original}$ | ECLG$^{FC}$ | ECLG$^{original}$ | ECLG$^{FC}$ | ECLG$^{original}$ | ECLG$^{FC}$ |
| number of papers | 12 | 6 | 38 | 18 | 19 | 10 |
| number of edges | 512 | 465 | 8898 | 7963 | 18094 | 15866 |
| number of nodes | 95 | 72 | 496 | 346 | 658 | 453 |
| mean papers per interaction ($MPI$) | 10.7 | 12.9 | 36 | 46 | 54.9 | 70 |
| mean interactions per paper ($MIP$) | 9.75 | 12 | 17 | 38.4 | 37 | 64.7 |
| path length ($APL_{avg}$) | 1.55 | 1.58 | 2.8 | 2.6 | 2.2 | 1.85 |
| maximum path length | 3 | 3 | 6 | 6 | 5 | 4 |
| minimum path length | 0 | 0 | 0 | 0 | 0 | 0 |
| clustering coefficient ($Coeff_{avg}$) | 0.977 | 0.979 | 0.98 | 0.99 | 0.99 | 0.99 |
| degree ($D_{avg}$) | 10.7 | 12.9 | 36 | 46 | 54.9 | 70 |
| number of clusters | 9 | 6 | 20 | 9 | 17 | 7 |
| neighborhood connectivity ($NC_{avg}$) | 11.29 | 13.3 | 37.8 | 46.6 | 56.8 | 70.6 |
| betweenness centrality ($BC_{avg}$) | 0.012 | 0.01 | 0.004 | 0.005 | 0.003 | 0.003 |
| closeness centrality ($CC_{avg}$) | 0.77 | 0.77 | 0.62 | 0.6 | 0.66 | 0.67 |
| radiality ($R_{avg}$) | 0.85 | 0.86 | 0.81 | 0.8 | 0.82 | 0.8 |
| edge betweenness ($EB_{avg}$) | 6.7 | 6.3 | 27.7 | 17.6 | 15.2 | 8 |
| frequency class ($FC_{avg}$) | 1.75 | 1.3 | 1.9 | 1.2 | 2 | 2 |

We also rank all events in the original set $\mathcal{E}$, based on the values of the node centrality metric with the highest $p_\varphi$, as selected by the PFI algorithm. Next, we choose the cut-off threshold for the centrality metric in order to select the most central nodes. We apply the threshold to determine a subset of $\mathcal{E}$, a new set $\mathcal{E}^{Central}$, that includes the most influential nodes, i.e., top ranked nodes according to the selected centrality metric. We refer to events in $\mathcal{E}^{Central}$, as *centrality candidate events*.

### IV. CASE STUDIES

In this section, we provide descriptions of three previously published computational models with $G(V, E)$ size varying from tens to hundreds of both nodes (in $V$) and edges (in $E$) (Table 1). We will use these models to test and evaluate our methodology under different conditions and scenarios. We chose these models as they were all carefully created manually and validated extensively against experimental results, while we used our proposed method to reconstruct the relevant pathways automatically.

The first model that we explore is the naïve T cell differentiation model that was introduced in [14] to help explain the differentiation of naïve T cells into the helper (Th) and the regulatory (Treg) phenotypes. The key markers that are commonly used to measure the outcomes of the naïve T cell differentiation are IL-2 and Foxp3, where Th (Treg) cells are characterized by the high (low) expression of IL-2 and low

(high) expression of Foxp3. In [14], the authors present a manually created logical model of the naïve T cell differentiation that recapitulated key experimental observations and generated several predictions. In [15], the authors presented a manually extended version of the original model from [14], and we will use this extended model as the golden model for this case study.

The second model that we explore is the T cell large granular lymphocyte (T-LGL) leukemia model that was presented in [16]. T-LGL leukemia is a disease characterized by an abnormal increase of T cells [16]. There is no curative therapy yet known for this disease. Hence, there is a crucial need to identify potential therapeutic targets. A manually created discrete dynamic model of the disease was published in [16], and this model will serve as the golden model for this case study.

The third model that we explore is a pancreatic cancer cell (PCC) model that was manually created, and discussed in [17]. The PCC model includes major signaling pathways, metabolic pathways, and the signals from the tumor microenvironment. The PCC receptors are included in the model such that the cell's response to external stimulations can be simulated over time. The authors also incorporated the hallmarks of cancer leading to suggestions of combinations of inhibitors as therapies. These hallmarks are represented in the PCC model as the processes of apoptosis, autophagy, cell cycle progression, inflammation, immune response, oxidative phosphorylation and proliferation. The model from [17] will serve as the golden model for this case study.

We created the EES of each case study as follows. For the T cell study, among 32 references cited by [15], we selected 12 most relevant papers, that is, papers in which T cell is mentioned together with one or more of the key elements of the model from [14]. Similarly, when creating the EES for the PCC study, we used the 19 papers cited in [17], as those papers provided evidence for the manually constructed PCC model in [17]. We used a different approach when assembling the EES for the T-LGL study. Instead of relying on the same literature that was used to manually build the published golden model, we created a search query "*T cell large granular lymphocyte (T-LGL) leukemia and proliferation and apoptosis*", and we used it as an input to the literature search engine PubMed [18]. From the papers that PubMed returned, we then selected the 38 papers that PubMed identified as "Best match". For each case study set of papers, we used an open source REACH [4] reading engine to extract events and create the corresponding EES. REACH is available online and can also be run through INDRA [6].

V. RESULTS AND DISCUSSION

We conducted several experiments using the three models described in Section IV. We explored how well the proposed methodology performs in various scenarios, small vs. large model and controlled vs. query-based EES.

A. *FC candidate events in three case studies*

For each case study, we create an ECLG. To identify the FC candidate events, we compute the *FC* value for all nodes

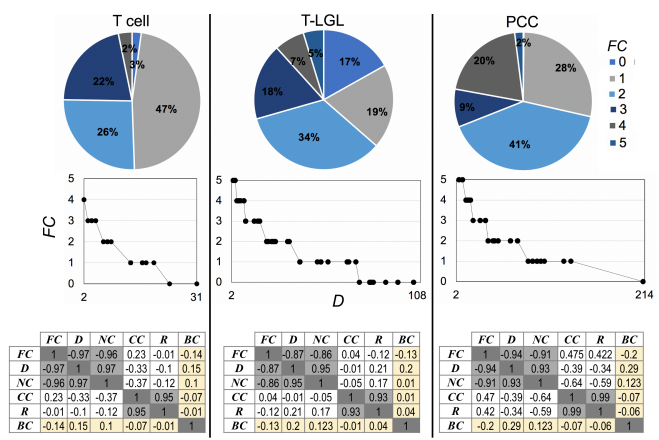

*Figure 2. Comparison of metric values for the three case studies, T cell, T-LGL, and PCC: (Top) Distribution of frequency class values within each EES. (Middle) FC vs. D values of all nodes in ECLG$^{original}$. (Bottom) Correlation coefficients between different metrics.*

(events) in the ECLG, according to equation (1). As described in Section III.A, events with *FC* = 0 are the most frequent ones and are considered to be strongly supported by literature, where multiple statements include them. We found that in all three case studies $FC_{avg} = 2$, and therefore, we will use this value as a threshold for removing less frequent events (i.e., all events with $FC > 2$) from the ECLG in each case study. We list in Table 1 the number of nodes and edges in the ECLG of each case study, both before (ECLG$^{original}$) and after (ECLG$^{FC}$) the removal of less frequent nodes, that is, the size of sets $\mathcal{E}$ and $\mathcal{C}$, and sets $\mathcal{E}^{FC}$ and $\mathcal{C}^{FC}$, respectively. We also show in Table 1 the centrality metric values for all ECLGs. To further compare and contrast the two versions of ECLG in each case study, ECLG$^{original}$ and ECLG$^{FC}$, we also show in Table 1 values for other commonly used graph metrics. As a reminder, the nodes in ECLG$^{FC}$ represent FC candidate events. As can be noticed in Table 1, not only there is a difference in size between sets $\mathcal{E}$ and $\mathcal{E}^{FC}$, and sets $\mathcal{C}$ and $\mathcal{C}^{FC}$, but also other graph parameters changed. For instance, the change in the average neighborhood connectivity value $NC_{avg}$ ranges from 2% for the T cell use case to 14% in the PCC use case. The distribution of *FC* values within the EES of each case study is illustrated with pie charts in Figure 2(Top). When the percentages in each *FC* category are averaged across the three case studies, the distribution of events with *FC*=(0, 1, 2, 3, 4, 5) is (7%, 31.3%, 33.6%, 16.3%, 9.6%, 2.3%), respectively. As expected, consistent across all three case studies, the number of distinct events that occur most frequently in literature (*FC*=0) is small compared to other categories. Interestingly, in all three case studies, the number of distinct events that do not have many occurrences in literature (*FC*=4 or *FC*=5) is also relatively small, while more than half of the total number of distinct events is in the higher occurrence frequency categories (*FC* = 2 or *FC* = 1).

B. *Evaluating centrality of FC candidate events*

We investigated the relationship between graph centrality metrics (described in Section III.C) and our proposed FC metric. To compare these metrics, we used the golden model and the corresponding EES from each case study. As the values in Table 1 suggest, for each case study, the removal of less frequent nodes leads to a denser graph, with strongly connected components, which is in agreement with both the increased $NC_{avg}$ value of the nodes and the high clustering coefficient. Moreover, the average node degree $D_{avg}$ in ECLG increased after the removal of less frequent nodes. This is due to the high (inverse) correlation between the D value and the event FC value, as shown in Figure 2(Middle). Furthermore, the less frequent events (with FC > 2), have the lowest D values. The most frequent events in literature, which are at the same time the nodes with higher D values, tend to have a greater ability to influence other ECLG nodes. The correlation coefficients of different metrics are illustrated in Figure 2(Bottom). The nodes having high D values also have FC=0. The strong correlation between FC and D values is also highlighted in the D distribution histogram for ECLG$^{original}$ and ECLG$^{FC}$ in Figure 3(Top).

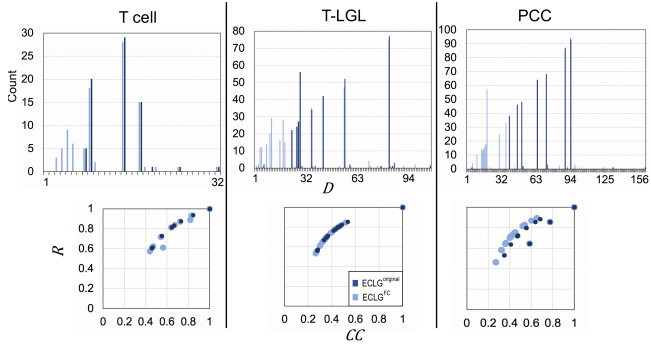
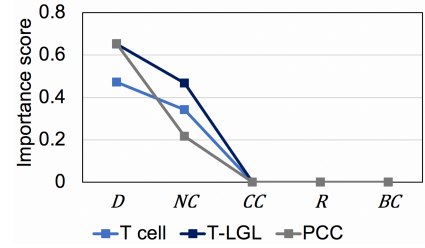

Figure 3. Exploration of graph characteristics for the three case studies, T cell model, T-LGL model, and PCC model: (Top) distribution of degree (D) values, (Bottom) Radiality (R) vs. closeness centrality (CC) of the ECLG network before and after the removal of less frequent events (ECLG$^{original}$ and ECLG$^{FC}$, respectively).

Figure 4. The importance score $p_\varphi$, computed as in Equation 2, for the centrality metrics D, NC, CC, R, and BC in the three case studies.

On the other hand, the closeness centrality CC values and the radiality R values of nodes do not seem to correlate with FC values. The R and CC values are highly correlated [24], that is, larger R and CC values indicate central position of a node in a graph, and this is also clearly seen in Figure 3 (Bottom). Interestingly, when we plotted R vs. CC of ECLG$^{original}$ and ECLG$^{FC}$, we found that in ECLG$^{FC}$ the relationship between the R and CC values takes almost a linear shape. The removal of the less frequent nodes took out the main outliers that existed in ECLG$^{original}$, while the R and CC values of the remaining nodes did not change much. The reason behind small changes in the remaining nodes is that some of the removed nodes were already separated from the main connected component of the ECLG$^{original}$ graph. Another centrality measure that is not correlated with the literature occurrence frequency is betweenness centrality BC. As listed in Table 1, the average BC value is approximately 0.01, both before and after the removal of less frequent nodes (i.e., in ECLG$^{original}$ and ECLG$^{FC}$, respectively).

As can be seen in the table in Figure 2(Bottom), there is high correlation between NC and FC values in ECLG nodes, which is also confirmed by the strong correlation between D and NC values, on one side, and the D and FC values, on the other. This is further confirmed using PFI to identify the centrality metric that contributed the most to classifying the EES – Figure 4 shows the importance score $p_\varphi$ of each centrality metric for all three case studies. We note that degree metric has the highest importance score, for all the case studies, the neighborhood connectivity centrality metric has the second highest and a non-zero importance score, whereas all the other centrality metrics have zero importance score.

### C. Evaluation of the proposed FC metric

For each case study, we compute the precision and recall of the FC metric and the degree centrality metric. To determine the precision and recall values, we consider the candidate events that are also present in the golden model as true positives or *true events*, and the remaining candidate events as false positives or *false events*. Similarly, the events that are in the golden model and were not selected as candidate events are false negatives and the events that are not in the golden model and were not selected as candidate events are true negatives. We will refer to the golden model events (also defined in Section II.A as directed element interactions) as *correct events*. Precision is the ratio between the number of true events and the sum of the number of true events and the number of false events, whereas recall is the ratio between the number of true events and the total number of correct events found in the EES (i.e., the sum of the number of true positive and the number of false negative events).

We show in Figure 5 the precision and recall results for the FC candidate events of the three case studies. For the T cell case, we achieved a precision of 0.44. This means that 56% of the FC candidate events are false positives (i.e., they are not in the golden model). On the other hand, for the T-LGL case, the event precision is 0.3, and in the PCC case, it is 0.25. While in the T-LGL and PCC studies more than half of the events and entities are false positives, it is important to note that these two studies have much larger EES, compared to the T cell study, and thus, have more candidate events. Moreover, the events that are in the golden models are not necessarily the only valid events, as there could be other events in literature that are also useful and important, and therefore, should be considered in model assembly. Our proposed methodology is able to uncover such events and suggest them as model extension candidates. For instance, in the PCC study, the events IL-6 → MMP, STAT3 → Twist, NF-κB → Bcl-2, NF-κB → Bcl-X L, STAT3 → Bcl-2 and STAT3 → Bcl-X L, AID —| P53 (where "→" represents positive regulation, and "—|" represents negative regulation) were identified as FC candidate events,

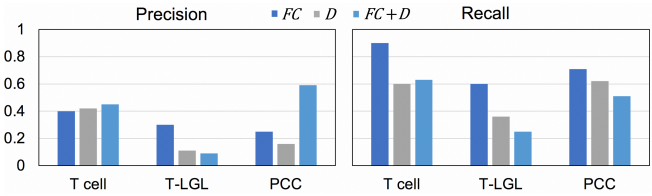

*Figure 5. Precision and recall values when compared to golden models for T cell, T-LGL and PCC use cases.*

and their correctness was approved by domain expert although they are not in the model. Examples of the evidence statements for those events, found in the REACH output, are: "IL-6 promotes MMP expression", "STAT3 mediated induction of Twist transcription", "the expression of the anti-apoptotic proteins Bcl-2 and Bcl-X L are promoted by both NF-κB and STAT3 and a novel mouse model of hepatocarcinogenesis triggered by AID causing deleterious p53 mutations". Therefore, the precision that we report in Figure 5 is likely smaller than the actual precision of our proposed method due to these additional important events that the method is able to uncover, and which are not in the golden model. To elaborate more on this, we conducted the following exercise for the PCC study. We used human judgement of candidate events, that is, based on domain expert's opinion, we labeled the candidate events as true or false positives. Interestingly, the domain expert identified additional 144 events as true positives (valid FC candidate events, but not in the golden model), besides the 151 true events (FC candidate events that are also in the golden model). When we changed the status of these 144 events from false positives to true positives, the precision increased to 0.65.

We used INDRA to compute a belief score for each selected event in the T-LGL study, where we used a query to search for papers instead of preselected list of papers. INDRA generated a belief score with value greater than or equal 0.7 (out of 1) for 22 additional events. When we changed the status of these events from false positives to true positives, this has increased the event precision from 0.3 to 0.4.

The recall values for our proposed method are much higher than the precision values (Figure 5). This demonstrates the ability of our methodology to identify useful and relevant events in a given EES. In particular, for the T cell study, the recall value is 1, as none of the correct events are missed, that is, there are zero false negatives. For PCC case, the recall value is 0.76. Finally, for the T-LGL case study, the recall is 0.70, i.e., our method missed approximately 30% of correct events. The lower recall in this study is due to removing a large number of events (41%) from the golden model to create the baseline model (Table 1), as well as using a large EES.

It is worth noting that the values of precision and recall are highly affected by the accuracy of machine readers. There are several errors that arise from machine reading output when extracting the events from published literature. For instance, a common error that we noticed in the PCC case study is related to the EGFR (epidermal growth factor receptor) protein. When the machine reader finds EGFR in a paper, it translates it into EGFrna which is not true and makes any event that contains the protein EGFR a false positive.

### D. Evaluation of the degree centrality metric

Since we found that the degree centrality metric highly correlates with our proposed FC metric, we were interested in exploring the difference between the set of centrality candidate events found using the degree metric and the set of FC candidate events found as described in Section III.D. We ranked the nodes (events) in the ECLG$^{original}$ in descending order of $D$ values, i.e., from the event with highest $D$ value to the event with lowest $D$ value. Similar to *FC*, we can set a threshold and remove the events with low $D$ values. Therefore, the top ranked nodes are the centrality candidate events to be evaluated.

For each case study, we computed the threshold as the average $D$ value, and we removed all the events that are below this value. The generated set is the set of centrality candidate events. As shown in Table 1, the average $D$ values are 10.7, 36 and 54.9 for T cell, T-LGL and PCC case studies, respectively. The number of events in the set $E^{Central}$, after the removal of the events with small $D$ values is 47 for the T cell case study, 215 for the T-LGL case study and 405 for the PCC case study. We show in Figure 5 the precision and recall of the centrality candidate events for the three case studies. For T cell study, the precision is 0.4 and the recall is 0.9. For the PCC study, the precision is very small, it is 0.16, and the recall is 0.6. Similarly, for the T-LGL study, the precision value is very small, it is 0.11 and the recall is 0.36. In all case studies, these values are smaller than the values reported for the FC metric. This is due to the centrality metric $D$ removing a subset of true events that were in the FC candidate event set. These results emphasize the fact that the FC metric is more accurate in extracting true events from the EES, compared to centrality metrics.

Finally, we conducted a two-step exercise, by first selecting the nodes (events) with $FC \leq 2$ from ECLG to form ECLG$^{FC}$. We then computed the average $D$ value of the nodes in ECLG$^{FC}$ as our new threshold and we removed all the nodes having $D$ value below this threshold. The number of events that we obtained are 48, 137 and 255 for the T cell, T-LGL and PCC cases, respectively. For those events, we compute the precision and recall as shown in Figure 5. For all use cases, the recall values are smaller than *FC* values of 0.63 for T cell, of 0.25 for T-LGL and of 0.51 for PCC. This means that more correct events were removed which reduces the recall values. However, for the PCC case, the precision value significantly increased to 0.5 since additional events that are likely false positives were removed. This suggests that using the two-step selection could be beneficial when EES is at the order of tens of thousands of interactions or larger.

### VI. CONCLUSION

Our proposed automated framework for rapid model assembly combines machine reading, the frequency class-based metric, and graph analysis. We compared the performance of our proposed frequency class-based metric to four common centrality metrics, and we also evaluated the usefulness of the centrality metrics when identifying the events that are highly supported in literature. Our results suggest that our proposed frequency class-based metric is most useful when the machine reading output has hundreds or thousands of events, while in the case of larger extracted events sets, the event selection can be further improved when the frequency class metric is used together with the degree centrality metric. Furthermore, our methodology automatically assembles models using the information published in literature within several seconds. As such, it facilitates information reuse and data reproducibility, and it could replace hundreds or thousands of manual experiments.